\documentclass[aps,pre,reprint,superscriptaddress]{revtex4-1}%
\usepackage{amsfonts}
\usepackage{amsmath}
\usepackage{amssymb}
\usepackage{graphicx}
\usepackage{color}
\usepackage{amsmath}
\usepackage{float}
\usepackage{hyperref}

\begin{document}
\title{Defects at grain boundaries: A coarse-grained, three-dimensional
description by the amplitude expansion of the phase-field crystal model}
\author{Marco Salvalaglio} \email{marco.salvalaglio@tu-dresden.de} 
\affiliation{Institute  of Scientific Computing,  Technische  Universit\"at  Dresden,  01062  Dresden,  Germany}
\author{Rainer Backofen} 
\affiliation{Institute  of Scientific Computing,  Technische  Universit\"at  Dresden,  01062  Dresden,  Germany}
\author{K. R. Elder}
\affiliation{Department of Physics, Oakland University, Rochester, 48309 Michigan, USA.}
\author{Axel Voigt}
\affiliation{Institute  of Scientific Computing,  Technische  Universit\"at
Dresden,  01062  Dresden,  Germany} \affiliation{Dresden Center for
Computational Materials Science (DCMS), TU Dresden, 01062 Dresden, Germany}

\begin{abstract}
We address a three-dimensional, coarse-grained description of dislocation networks at grain boundaries between rotated crystals. The so-called amplitude expansion of the phase-field crystal model is exploited with the aid of finite element method calculations. This approach allows for the description of microscopic features, such as dislocations, while simultaneously being able to describe length scales that are orders of magnitude larger than the lattice spacing. Moreover, it allows for the direct description of extended defects by means of a scalar order parameter. The versatility of this framework is shown by considering both fcc and bcc lattice symmetries and different rotation axes. First, the specific case of planar, twist grain boundaries is illustrated. The details of the method are reported and the consistency of the
results with literature is discussed. Then, the dislocation networks forming at
the interface between a spherical, rotated crystal embedded in an unrotated
crystalline structure, are shown. Although explicitly accounting for
dislocations which lead to an anisotropic shrinkage of the rotated grain, the
extension of the spherical grain boundary is found to decrease linearly over
time in agreement with the classical theory of grain growth and recent
atomistic investigations. It is shown that the results obtained for a system with bcc 
symmetry agree very well with existing results, validating the methodology. 
Furthermore, fully original results are shown for fcc lattice
symmetry, revealing the generality of the reported observations.
\end{abstract}

\maketitle

\section{Introduction}
\label{sec:introduction}

Grain boundaries (GBs) are interfaces between crystal grains having different
orientations. They consist of extended, two-dimensional defects and determine
several features of polycrystalline materials such as mechanical 
and ferromagnetic properties as well as thermal and electrical 
conductivity \cite{Sutton1995}. The migration of GBs
determines the microstructure evolution in polycrystalline materials
\cite{Demirel2003} and consequently it is crucial to 
develop an in-depth understanding of such motion for controlling
and tuning material properties \cite{Randle2004}.

The description of GB morphologies and dynamics
is a typical mesoscale problem \cite{Rollett2015}. Indeed, networks of
one-dimensional, extended defects, i.e.
dislocations, form at the interfaces between crystals with different
orientations \cite{Sutton1995}. Their nature, motion and reaction, in turn,
strictly depends on microscopic features such as the atom packing along
crystallographic planes or the concentration of vacancies \cite{hl82}.
The importance of GBs and their influence on microstructural behavior in a wide range of 
polycrystalline materials, from metals and semiconductors
\cite{Merkle1990,Grovenor1985} to ionic and organic crystals
\cite{Duffy1986,Haruta2012}, make them a central topic for material science
investigations. 
Indeed, the study of GBs fostered the development of several
experimental and theoretical techniques working at different length scales
\cite{Rollett2015}.

Regarding theory and modeling, many of the available methods from the atomic to
the continuum lengthscale has been adopted in order to provide both a deep
understanding of experiments and insights on the GB behaviors under ideal
conditions. Continuum modeling shed first light on the basics of GB energetics,
motion, and morphology \cite{Read1950,Kobayashi2000}. Macroscopic simulations of
microstructures evolution has been used to address the dynamics of GBs as a
continuous interface between grains, e.g., by phase- and multiphase-field
approaches \cite{Steinbach1999,Kim2006,Kim2014} or by continuum mechanics
simulations accounting for plasticity \cite{Jung2017}.  Microscopic methods
such as molecular dynamics (MD) or Monte-Carlo approaches have been used to
provide atomic-level insights on GBs and parameters for mesoscale or
macroscopic modeling
\cite{Farkas2000,Mishin2010,Mendelev2013,Anderson1989,Yu2003}. The so-called
phase-field crystal (PFC) method \cite{Elder2002,Elder2004,Emmerich2012}
provides a description of the atomic probability 
density on the diffusive timescale and filters out 
the dynamics of vibrations. This approach has been extensively
used to provide long-timescale descriptions of GBs and
dislocation dynamics \cite{Adland2013,Berry2014,Yamanaka2017}. 

While each of these methods is a viable tool for the investigation of GBs,
they have limitations. For instance, MD is restricted to relatively short timescales limiting the application to long
timescale processes, PFC requires a fine spatial discretization allowing for
the simulation of relatively small systems and many continuum approaches are 
missing the essential physical properties of GBs. Therefore, multiscale methods or coarse-grained approaches,
bridging the gap between the range of applicability of these techniques, would
be highly desirable to provide general information on polycrystalline
materials.

The so-called amplitude expansion of the PFC model (APFC) offers a
coarse-grained description able to account for atomistic details at
lengthscales typical of classical phase-field (or in general continuum)
approaches \cite{Goldenfeld2005,Athreya2006,GoldenfeldJSP2006}. The
periodic atomic probability density described by PFC is accounted for by means
of the minimum set of Fourier modes or plane waves required by a given
crystal-lattice symmetry. Then the evolution of the complex amplitudes of such waves,
$\eta_j$, is described under the approximation of slowly varying amplitude
functions, i.e., varying on a larger lengthscale than the lattice spacing.
Complex amplitude functions allow for describing rotations and deformations of
the lattice structure. In addition, by means of $\eta_j$'s it is also possible to
compute a scalar order parameter (referred to as $A^2$ in the following) that
is maximum within the crystal, decreases at defects and at crystal-melt or
ordered-disordered interfaces, and vanishes for disordered-liquid phases, so
that it can be thought of as being related to the order parameter 
that enters standard phase-field
approaches.
Although the APFC approach does not provide an accurate description 
of the atomic rearrangement at dislocation cores,
it is known to give good coarse-grained approximations of PFC for small
deformations or tilts and has been already adopted to investigate GBs in two
dimensions \cite{Adland2013,Geslin2015,Xu2016,Huter2017}. Moreover, the
original model has been extended to binary systems and to body-centered cubic
(bcc) and face-centered cubic (fcc) symmetries
\cite{ElderPRE2010,ElderJPCM2010}. The possibility to tune interface and defect
energies has been also recently shown \cite{SalvalaglioAPFC2017}. Except for
some proof of concepts, no three-dimensional (3D) simulations were provided.
However, in Ref.~\cite{SalvalaglioAPFC2017} a unified description of different
lattice symmetries and a numerical scheme able to cope with 3D systems were discussed, 
paving the way to advanced simulations of material properties.

In this paper, by exploiting the framework reported in
Ref.~\cite{SalvalaglioAPFC2017}, we address the 3D description of defect
structures forming at the interface between rotated crystals by APFC. A
coarse-grained description of dislocation networks, achieved without explicitly
considering the atoms of the crystal lattice, is then provided. It is proved to
account for the specific lattice symmetry, the rotation axis, and the
geometry of the interface between crystals. The final goal of this paper is to
illustrate defect morphologies forming between a spherical,
rotated grain in an unrotated lattice and their evolution. Indeed, this is a prototypical system
where all the possible orientations of a GB with respect to the rotation axis
are present. Moreover, it evolves until the disappearance of the rotated grain
and can be compared with classical continuum theory \cite{Doherty1997} and
atomistic calculations \cite{Yamanaka2017}.

To this purpose, we first tackle the modeling of planar, twist GBs. This
allows us to assess the general approach and show the versatility of the method
in describing different symmetries and GB orientations. Moreover, we show how
the periodicity of the defect structures is related to single Fourier modes and
we illustrate how to exploit this connection to provide efficient simulations.
The morphologies of the defect structures are found to correspond to the
networks minimizing the GB energy as reported in the literature. The effect of
different twist angles on the features of dislocation networks as well as on
the GB energies are also discussed. 

Afterwards, the closed dislocation networks forming spherical GBs due to
rotated crystal inclusions are addressed. Their morphologies are discussed also
pointing out the similarities with the defect networks of planar, twist GBs
reported previously.  The anisotropic shrinkage of rotated spherical grains and
the connection with standard continuum theory are then discussed. 
Through the entire paper both fcc and bcc lattice symmetries are considered,
proving the versatility of the method and the generality of the reported
observations. It is worth mentioning that similar investigations adopting
standard PFC were recently provided in Ref.~\cite{Yamanaka2017} for bcc lattice symmetry only. This agreement further validates the coarse-grained approach considered here. Moreover, 
new results are shown for fcc symmetry and allow for the identification of general features
occurring during the shrinkage of rotated grains.

The paper is organized as follows. In Sec.~\ref{sec:model} the APFC approach is briefly illustrated, while the modeling and computational details developed for the investigations reported in the following sections are summarized in Sec.~\ref{sec:simdetails}. Results concerning planar, twist GBs are reported in Sec.~\ref{sec:planar}. The morphologies and the shrinkage of spherical rotated crystals are shown and discussed in Sec.~\ref{sec:spherical}. Conclusion and remarks are summarized in Sec.~\ref{sec:conclusions}.

\section{APFC Model}
\label{sec:model}

The standard PFC approach is based on the definition of a free energy $F_n$ as function of the local atomic density $n$ \cite{Elder2007}:
\begin{equation}
F_n=\int_{\Omega} \left[\frac{\Delta B_0}{2}n^2+\frac{B^x_0}{2} n(1+\nabla^2)^2n 
-\frac{t}{3}n^3+\frac{v}{4}n^4 \right]d\mathbf{r},
\label{eq:F_PFC}
\end{equation}
where $\Delta B_0$, $B_0^x$, $v$, and $t$ are positive parameters controlling the properties of the system \cite{Elder2007}. The coarse grained approach considered in this work consists in the so-called amplitude expansion \cite{Goldenfeld2005,Athreya2006,GoldenfeldJSP2006} of the PFC model (APFC), where the atomic density is expressed as
\begin{equation}
n=n_0+\sum_{j=1}^N \left[ \eta_j(\mathbf{x},t) e^{i\mathbf{k}_j \cdot
\mathbf{x}}+ \eta_j^*(\mathbf{x},t) e^{-i\mathbf{k}_j \cdot \mathbf{x}}\right],
\label{eq:density}
\end{equation}
with $\mathbf{k}_j$ the $N$ nonzero reciprocal-lattice vectors reproducing a specific lattice symmetry and $n_0$ the average density which can be set to zero without loss of generality \cite{ElderPRE2010}. For bcc lattice symmetry $N=6$ and the $\mathbf{k}_j/k_0$ vectors read:
$\mathbf{k}_1$=$\left(1,1,0\right)$, 
$\mathbf{k}_2$=$\left(1,0,1\right)$,
$\mathbf{k}_3$=$\left(0,1,1\right)$, 
$\mathbf{k}_4$=$\left(0,1,\bar{1}\right)$, 
$\mathbf{k}_5$=$\left(1,\bar{1},0\right)$, 
$\mathbf{k}_6$=$\left(\bar{1},0,1\right)$, with $k_0=\sqrt{2}/2$.
For fcc lattice symmetry $N=7$ and the $\mathbf{k}_j/k_0$ vectors read:
$\mathbf{k}_1$=$\left(\bar{1},1,1\right)$, 
$\mathbf{k}_2$=$\left(1,\bar{1},1\right)$, 
$\mathbf{k}_3$=$\left(1,1,\bar{1}\right)$, 
$\mathbf{k}_4$=$\left(\bar{1},\bar{1},\bar{1}\right)$, 
$\mathbf{k}_5$=$\left(2,0,0\right)$,
$\mathbf{k}_6$=$\left(0,2,0\right)$,
$\mathbf{k}_7$=$\left(0,0,2\right)$, with $k_0=\sqrt{3}/3$. Further details can be found in Refs.~\cite{ElderPRE2010,SalvalaglioAPFC2017}.

In the APFC approach the state and the evolution of the system are described by
the amplitude functions $\eta_j$, which are complex functions. The free energy
functional of the APFC approach is then written in terms of $\eta_j$'s.
Assuming that these functions vary on lengthscales much larger than the atomic
spacing, the free energy reads \begin{equation}
\begin{split}
F=\int_{\Omega} &\bigg[\frac{\Delta B_0}{2}A^2+\frac{3v}{4}A^4 +\sum_{j=1}^N
\left ( B_0^x |\mathcal{G}_j \eta_j|^2-\frac{3v}{2}|\eta_j|^4 \right )
\\ & +f^{\rm s}(\{\eta_j\},\{\eta^*_j\}) \bigg]  d \mathbf{r}, \end{split}
\label{eq:energyamplitude}
\end{equation}
where $\mathcal{G}_j\equiv \nabla^2+2i\mathbf{k}_j \cdot \nabla$ and 
\begin{equation}
A^2\equiv 2\sum_{j=1}^N |\eta_j|^2.
\label{eq:A2}
\end{equation}
The parameters are set as follows: $B^x=0.98$, $v=1/3$, $t=1/2$, and $\Delta
B=0.02$ \cite{ElderPRE2010,SalvalaglioAPFC2017}. $A^2$ describes the phases of the system; it is constant for a
bulk crystal, decreases at solid-liquid interfaces or at defects, and
vanishes in disordered or liquid phases. $f^{\rm
s}(\{\eta_j\},\{\eta_j^*\})$ are complex polynomials generated by the cubic 
and quartic terms in the free energy $F_n$, whose exact form depends on 
crystal symmetry as reported in Ref.~\cite{SalvalaglioAPFC2017}. The
evolution laws for $\eta_j$'s read 
\begin{equation}
\frac{\partial \eta_j}{\partial t} =-|\mathbf{k}_j|^2 \frac{\delta F}{\delta \eta_j^*},
\label{eq:amplitudetime}
\end{equation}
with
\begin{equation}
\begin{split}
\frac{\delta F}{\delta \eta_j^*}=& \left[
\Delta B_0 + B_0^x\mathcal{G}_j^2 + 3v \left(A^2-|\eta_j |^2\right)\right]\eta_j \\ 
& + \frac{\delta f^s(\{\eta_j\},\{\eta^*_j\})}{\delta \eta_j^*} .
\end{split}
\label{eq:amptimefuncder}
\end{equation}

By means of $\eta_j$'s, it is possible to account for distortion of the lattice. 
Indeed they can be generally written in the polar form as
\begin{equation}
\eta_j (\mathbf{r}) = \bar{\eta}_j e^{i \varphi_j(\mathbf{r}) \cdot \mathbf{r}}.
\label{eq:amprot}
\end{equation}
$\bar{\eta}_j$ can be determined by the minimization of the free energy
functional assuming equal and constant amplitudes for each different
$\mathbf{k}_j$ length, i.e., by considering a relaxed crystal
\cite{ElderPRE2010,SalvalaglioAPFC2017}. The phase term $\varphi_j(\mathbf{r})$
can instead account for displacements of the lattice with respect to a
reference state. In this paper, we are interested in describing GBs
between rotated, three-dimensional crystals. These systems can be described by
setting a phase term that quantifies the difference between the
reciprocal-lattice vectors in the rotated system $\mathbf{k}_j^{\theta}$ and
the original ones $\mathbf{k}_j$. This can be generally written as
\begin{equation}
\varphi_j(\mathbf{r})=\delta\mathbf{k}_{j}(\theta)\Theta(d(\mathbf{r})), 
\label{eq:theta}
\end{equation}
with $\Theta$ the Heaviside function, $d(\mathbf{r})$ the signed distance from the interface between a rotated ($d(\mathbf{r})>0$) and an unrotated ($d(\mathbf{r})<0$) crystal, and $\delta\mathbf{k}_{j}(\theta)=\mathbf{k}_{j}^{\theta}-\mathbf{k}_{j}$. The latter, for a rotation of an angle $\theta$ about the $\hat{\mathbf{x}}$-axis, reads
\begin{equation}
\begin{split}
\delta\mathbf{k}_{j}(\theta) = & \left[k^y_{j} (\cos\theta -1) - k^z_{j}
\sin\theta\right]\hat{\mathbf{y}} \\ &+ \left[k^y_{j} \sin\theta  + k^z_{j}
(\cos\theta-1)\right]\hat{\mathbf{z}} .
\end{split}
\label{eq:krot}
\end{equation}

Equations~\eqref{eq:amprot}, \eqref{eq:theta}, and \eqref{eq:krot} can then be used in order to specify specific rotations as a function of coordinates. They will be adopted to set the initial condition for simulations. Notice that $\delta\mathbf{k}_{j}(\theta)$ directly defines the wavelength of the amplitude oscillations as $\lambda_i^j=2\pi/[\delta\mathbf{k}_{j}(\theta)]_i$ with $i=x,y,z$.

\subsection{Simulation and modeling details}
\label{sec:simdetails}

The equations defined in \eqref{eq:amplitudetime} and \eqref{eq:amptimefuncder}
are solved by a finite element approach exploiting the toolbox AMDiS \cite{Vey2007,Witkowski2015}. We used a semi-implicit time discretization scheme for each amplitude. The resulting $N$
systems of second-order partial differential equations are then coupled by evaluating explicitly the terms originating from
$f^{\rm s}(\{\eta_j\},\{\eta^*_j\})$ which depend on more than one $\eta_j$.
Mesh refinement and adaptivity are used in order to ensure the proper
spatial discretization at defects. Conversely, a coarse mesh is sufficient to describe the
long wavelength oscillations of amplitudes as well as constant values for
unrotated crystals. The details of the numerical method, as well as the mesh-refinement criterion, are reported in
Ref.~\cite{SalvalaglioAPFC2017}. 

\begin{figure}%[H]
\center
    \includegraphics[width=\linewidth]{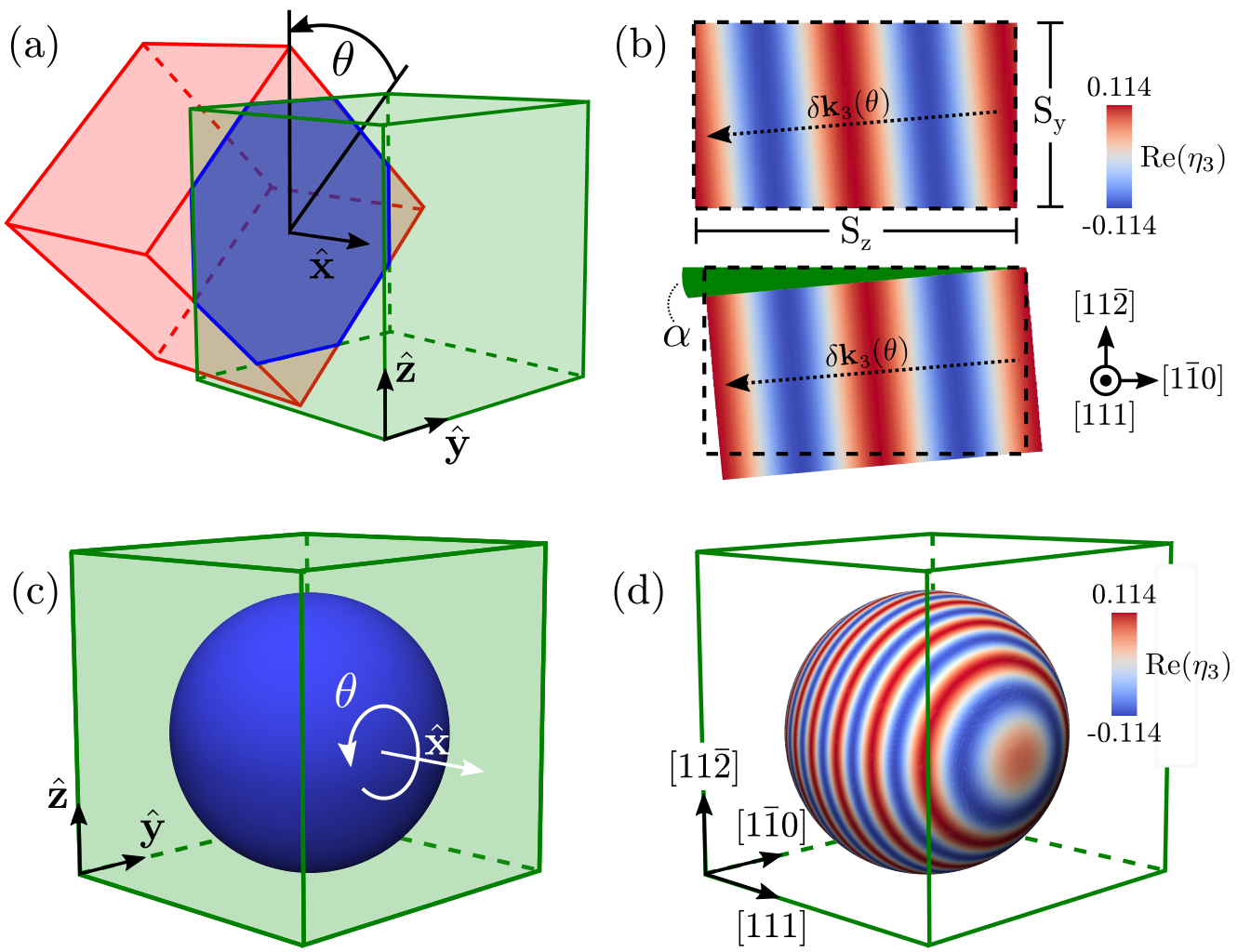} 
    \caption{Details about the simulation setup. (a) Schematics of a planar, twist
GB. (b) yz cross-section for a crystal with fcc symmetry rotated about the [111] with
$\theta=10^\circ$. The two panels show Re$(\eta_3)$ with (below) and without
(above) the rotation of the angle $\alpha$, chosen to have $\delta
\mathbf{k}_3(\theta)$ aligned to the boundaries. This allows for the smallest
$S_y$ and $S_z$ to reproduce the dislocation network at the GB. (c) Schematics
of a rotated, spherical crystal embedded in an unrotated crystal. (d)
Illustration of the Re$(\eta_3)$ for the initial condition leading to spherical
GBs with $\theta=10^\circ$.} \label{fig:figure1}
\end{figure}

The $\eta_j$'s are initialized by Eqs.~\eqref{eq:amprot}, \eqref{eq:theta}, and
\eqref{eq:krot}, i.e., a rotation of crystals about the $\hat{\mathbf{x}}$-axis
is set as illustrated in Fig.~\ref{fig:figure1}(a). Periodic boundary
conditions (PBCs) are set for all the amplitudes. In order to simulate planar,
twist GBs we set $d(\mathbf{r}) \equiv x $, having then a
grain boundary with normal $\hat{\mathbf{x}}$ at $x=0$. Due to PBCs a second GB
is expected, that is shared between the periodic boundaries with normal along
$\hat{\mathbf{x}}$. In the following only the GB at $x=0$ will be shown.

We consider two orientations for the twist GBs. In particular, we consider
rotations about $[ 111 ]$ and $[ 110 ]$ directions. For the former we set
$\hat{\mathbf{x}}=$[111], $\hat{\mathbf{y}}=$[1$\bar{1}$0], and
$\hat{\mathbf{z}}=$[11$\bar{2}$]. For the latter we set
$\hat{\mathbf{x}}=$[110], $\hat{\mathbf{y}}=$[$\bar{1}$10], and
$\hat{\mathbf{z}}=$[001]. Notice that some of the original $\mathbf{k}_j$ are
aligned with the $\hat{\mathbf{y}}$ and $\hat{\mathbf{z}}$ axes, so that a
small angle rotation would produce small components of $\delta
\mathbf{k}_j(\theta)$ and, in turn, large wavelengths $\lambda_i^j$. In order
to use PBCs, the size of the simulation domain along $\hat{\mathbf{y}}$ or
$\hat{\mathbf{z}}$ should then be a multiple of all the $\lambda^j_i$ with
$i=y,z$, respectively, and $j=1,...,N$. This would lead to very large
computational domains and unfeasible simulations. To overcome this issue, a
rotation of the simulation cell of an angle $\alpha$, chosen in order to have the smallest
of the $\delta \mathbf{k}_j(\theta)$ vector aligned with the boundaries, is
then considered. This allows us to set the smallest sizes of the simulation
domain, hereinafter referred to as $S_i$ with $i=x,y,z$. In particular this holds true for domain sizes
along the new $\hat{\mathbf{y}}$ and $\hat{\mathbf{z}}$ directions (now rotated
by $\alpha$), namely $S_y$ and $S_z$. The size $S_x$ is arbitrary as it sets
the distance between planar GBs. It is set to $50\pi$. An example of this
argument is shown in Fig.~\ref{fig:figure1}(b) for an fcc lattice. Two yz cross-sections of the
simulation domain adopted to reproduce a (111) planar twist GB with
$\theta=10^\circ$ are shown, with (below) and without (above) the rotation of
the simulation domain of the angle $\alpha$ about the [111] direction. Notice that
without such a rotation the values of Re$(\eta_3)$ do not satisfy PBCs. To recover these conditions, a much larger simulation domain is required. Conversely, having $\delta \mathbf{k}_3$ parallel to the sides allows
for using the domain with $S_i$ as in Fig.~\ref{fig:figure1}(b) and PBCs. For the sake
of simplicity, the dislocation networks will be shown upon a rotation of
$-\alpha$ in order to have the boundaries aligned with the original frame of
reference.

In order to simulate spherical, rotated crystal and, in turn, spherical GBs as shown in
Fig.~\ref{fig:figure1}(c) we set $d(\mathbf{r}) \equiv R-|\mathbf{r}| $. It corresponds to a rotated crystal inclusion in an
unrotated crystal. This allows us to set the domain size independently of the
rotation of the crystals as $\eta_j$ will not oscillate at the boundaries. In
particular we set $R=60\pi$ and $S_i=140\pi$. The same orientations
and frames of reference as for planar GBs are used. Figure~\ref{fig:figure1}(d)
shows Re$(\eta_3)$ [as in Fig.~\ref{fig:figure1}(b)] for fcc lattice symmetry
within the embedded crystal, rotated about the $[111]$ direction.

\begin{figure}[H]
\center
    \includegraphics[width=\linewidth]{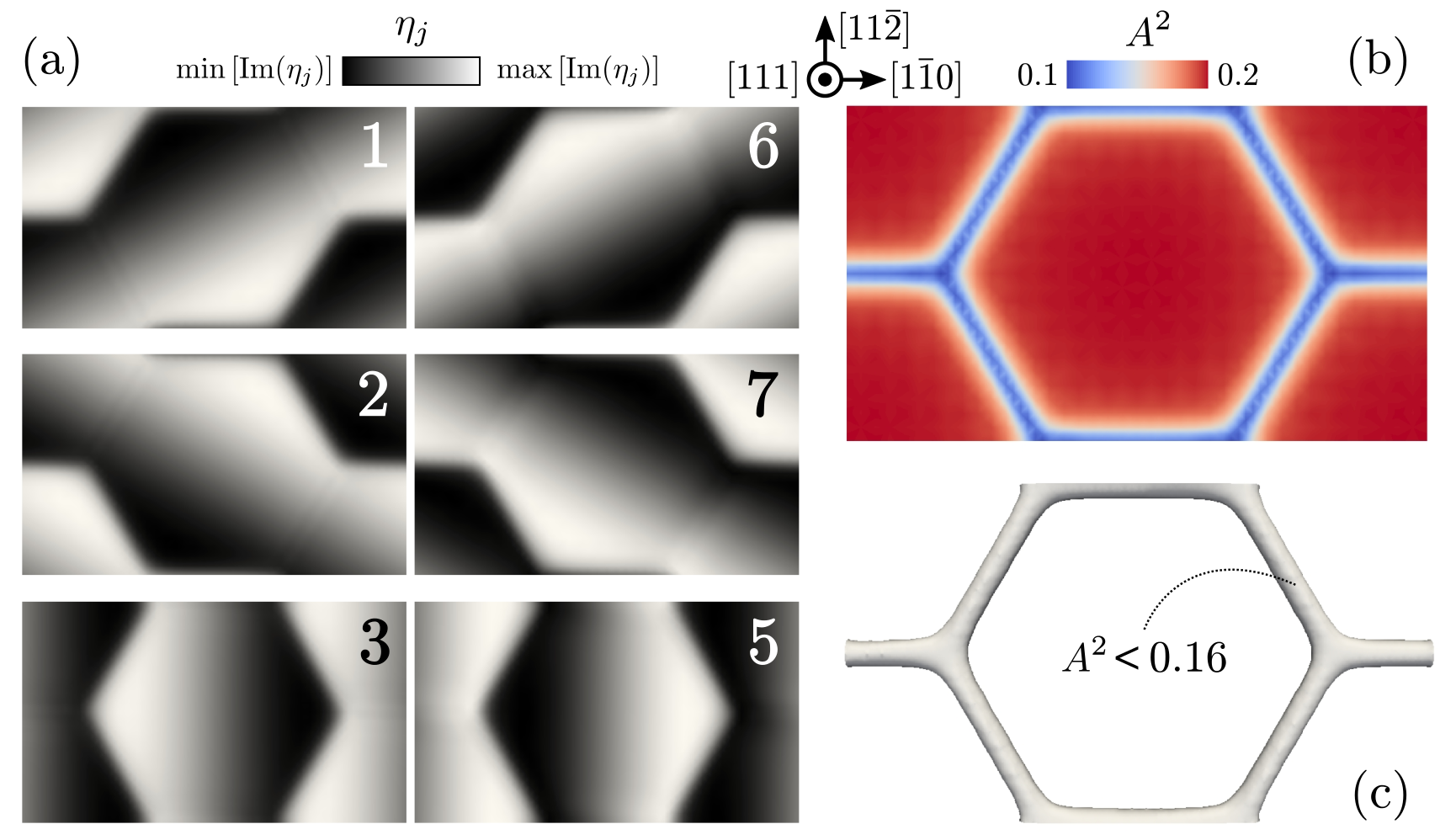} 
    \caption{Equilibrium condition for a (111) twist GB with fcc symmetry ($\theta=10^\circ$), reproduced by the APFC model. (a) Im$(\eta_j)$ [Im$(\eta_4) \sim 0$ not shown]. Numbers correspond to the value of the index in $\mathbf{k}_j$. (b) $A^2$ from Eq.~\eqref{eq:A2} computed from the amplitude functions illustrated in panel (a). (c) Dislocation network identified as a region where $A^2<0.8\max(A^2)$.}
    \label{fig:figure2}
\end{figure}

Figure~\ref{fig:figure2} shows various 
aspects of the APFC simulations. The aforementioned procedure
illustrated in Fig.~\ref{fig:figure1}(a) and Figure~\ref{fig:figure1}(b) is
used. Moreover, the relaxation of the initial condition is evaluated by
integrating Eqs.~\eqref{eq:amplitudetime} and \eqref{eq:amptimefuncder} as in
Ref.~\cite{SalvalaglioAPFC2017}. Fig.~\ref{fig:figure2}(a) illustrates the
imaginary part of amplitudes $\eta_j$ at the GB (i.e., the yz plane of the 3D
domain at $x=0$) with $\theta=10^\circ$ (after the $-\alpha$ rotation), with
the corresponding value of the index $j$ reported on each panel. Real parts
(not shown) exhibit similar features. The solutions for $\eta_j$ illustrated in
Fig.~\ref{fig:figure2}(a) represent the amplitude of the Fourier modes to be
used in Eq.~\eqref{eq:density} at the equilibrium. They directly reflect the
features and symmetries of $\mathbf{k}_j$ vectors. Panels on the same row refer
to $\eta_j$ which have antiparallel projections of $\mathbf{k}_j$ on the
yz plane. Indeed, the same features of the resulting fields are obtained with
an opposite sign only. $\mathbf{k}_4$ is aligned to the rotation axis, so the
components of $\delta\mathbf{k}_4$ are equal to zero, the initial condition for
$\eta_4$ consists of a real function with $\eta_4=\bar{\eta}_4$, and Im$(\eta_4)
\sim 0$, so that it is not shown in Fig.~\ref{fig:figure2}(a). 

Figure~\ref{fig:figure2} also illustrates the steps adopted in order to recognize
and show defects by means of solutions in terms of $\eta_j$ fields. The order
parameter $A^2$ is computed from the amplitudes by Eq.~\eqref{eq:A2}. As shown
in Fig.~\ref{fig:figure2}(b), $A^2$ is constant except for localized regions.
These regions correspond to defects of the crystal
lattice \cite{Goldenfeld2005,Athreya2006,GoldenfeldJSP2006} and in particular
to the network of dislocations defining the GBs that will be discussed in the
following. Therefore, a coarse-grained description is achieved, directly
accounting for defects and thus describing the crystal lattice in an effective
way. The results of simulations will be shown focusing on these dislocation
networks, as illustrated in Fig.~\ref{fig:figure2}(c). In this figure, the
three-dimensional region where $A^2<0.8\max(A^2)$ is reported. Notice that
different features of the resulting dislocation network as in
Fig.~\ref{fig:figure2}(c) are shared between some of the $\eta_j$'s, revealing
their contribution in determining the morphologies of GBs. Moreover, 
the system size ($S_i$) allows for simulating exactly one unit cell of the resulting
dislocation network.
 
\section{Planar Grain Boundaries}
\label{sec:planar}

\begin{figure}%[H]
\center
    \includegraphics[width=\linewidth]{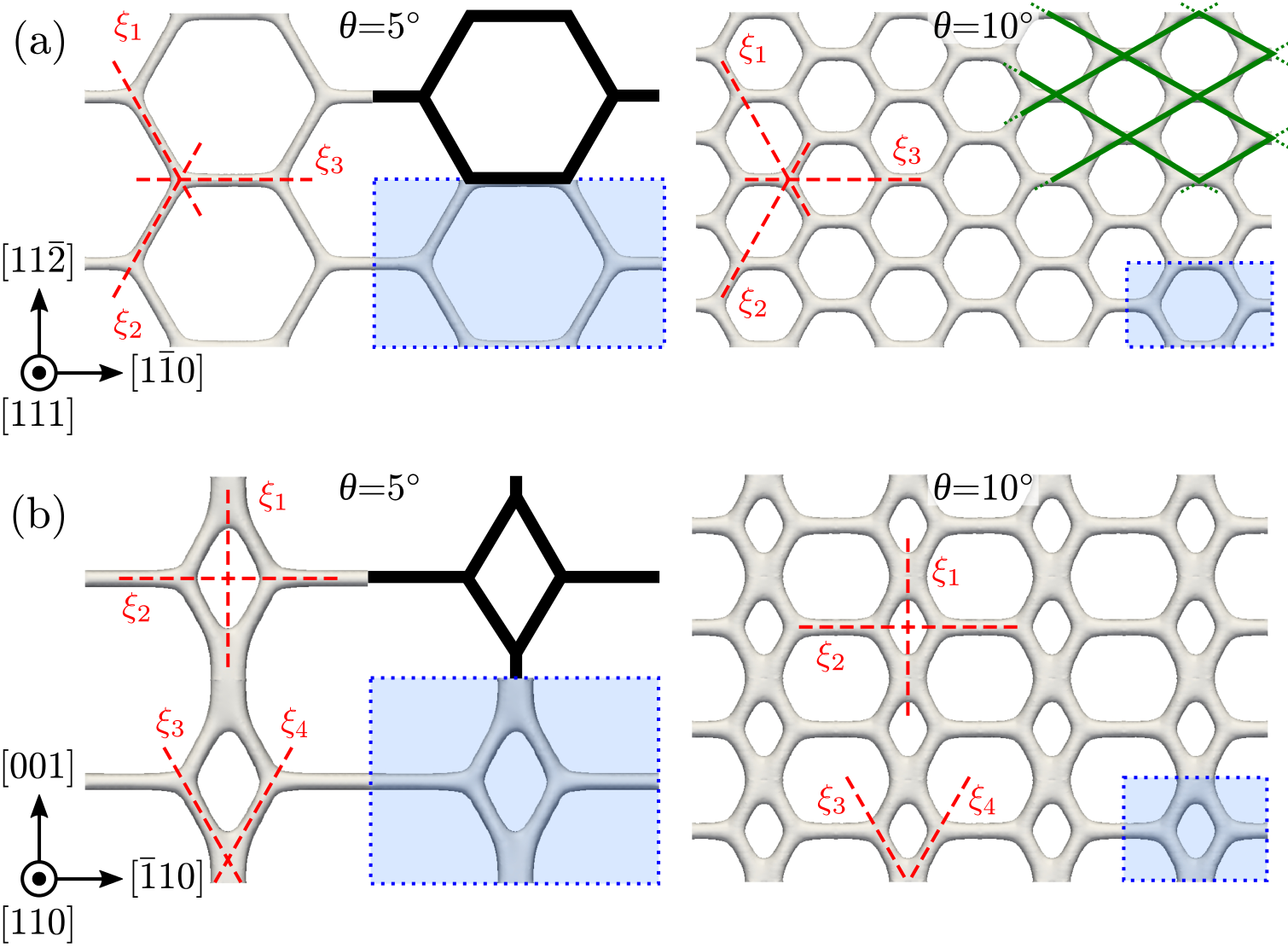} \caption{Planar twist-GBs,
fcc symmetry. Dislocation networks corresponding to $\theta=5^\circ$ (left) and
$\theta=10^\circ$ (right) are reported. Blue dotted line and shaded areas
illustrate the simulation domain for each case. Dislocation directions
$\boldsymbol{\xi}_i$ are highlighted by red dashed lines. Solid black lines
represent a schematics of the resulting dislocation networks. (a) (111) twist GB. Green solid lines in the $\theta=10^\circ$ panel illustrate the minimal set of dislocations accommodating the misorientation, still not minimizing the energy. 
(b) (110) twist GB.}
\label{fig:figure3}
\end{figure}

\begin{figure}%[H]
\center
    \includegraphics[width=\linewidth]{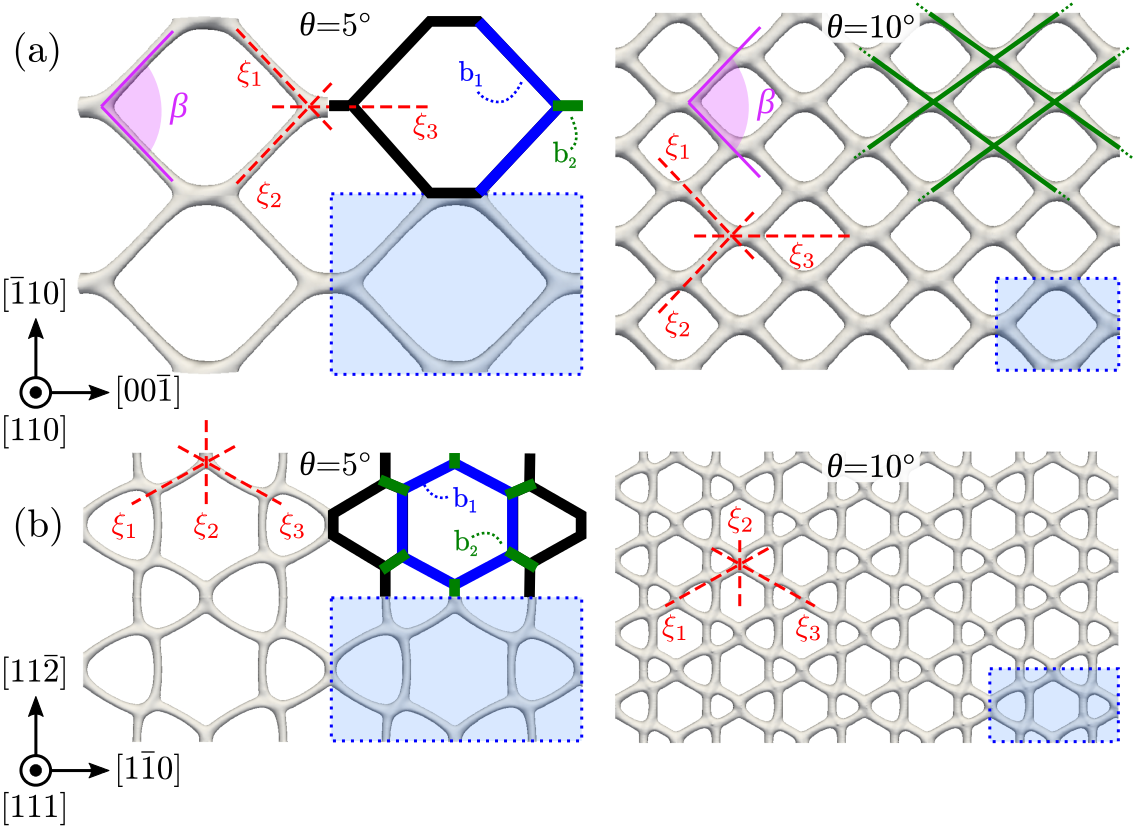} 
    \caption{Planar twist-GBs, bcc symmetry. Dislocation networks corresponding
to $\theta=5^\circ$ (left) and $\theta=10^\circ$ (right) are shown. Blue dotted
line and shaded areas illustrated the simulation domain for each case.
Dislocation directions $\boldsymbol{\xi}_i$ are highlighted by red dashed lines.
Solid lines represent the schematics of the resulting dislocation networks.
They illustrate also the two different expected Burgers vectors:
$\mathbf{b}_1=a/2\langle 111 \rangle$ (blue) and $\mathbf{b}_2=a\langle 100
\rangle$ (green). (a) (110) twist GB. The angle between $\boldsymbol{\xi}_1$
and $\boldsymbol{\xi}_2$, $\beta$, is also illustrate. Green solid lines in the $\theta=10^\circ$ panel illustrate the minimal set
of dislocations accommodating the misorientation, still not minimizing the energy. (b) (111) twist GB.} 
\label{fig:figure4}
\end{figure}

In this section we show the results for dislocation networks at twist GBs simulated by APFC. The procedure described in Sec.~\ref{sec:simdetails}
is used to generate the initial configurations as shown in Fig.~\ref{fig:figure2}. 
Figures~\ref{fig:figure3} and \ref{fig:figure4} illustrate the results in terms
of dislocation networks at GBs obtained for different orientations and lattice
symmetries. Two twist angles $\theta$, namely $\theta=5^\circ$ and
$\theta=10^\circ$, are considered. Notice that angles can be chosen here in a
continuous fashion as only long wavelength amplitudes have to be defined. The
comparison between simulations with different $\theta$ shows that the
dislocation networks at the GBs are preserved from a qualitative point of view,
while the dislocation density increases when increasing the twist angle.
Details concerning the dislocation network obtained for each specific symmetry
and orientation will be discussed in the following. The size of the simulation
cell in the plane perpendicular to the rotation axis is illustrated by a blue
dotted line and a shaded area. Notice that the smaller the tilt the larger the
simulation cell is, in agreement with the definition of $\lambda_i^j$, so that
an intrinsic limit in simulation feasibility exists for very small angles.
Solid thick lines illustrate the feature of the unit cell for dislocation
networks, which have a size corresponding to the simulation domain. The
simulated networks are repeated four times ($2\times 2$) for $\theta=5^\circ$  and 16 times ($4 \times 4$) for
$\theta=10^\circ$ to illustrate the resulting dislocation pattern. As a check,
the same dislocation networks have been obtained by simulating the entire
domains reported in Figs.~\ref{fig:figure3} and \ref{fig:figure4}. Notice that a
rotation of $-\alpha$ has been performed on the reported dislocation networks
(see also Sec.~\ref{sec:simdetails}). 

Let us first focus on the twist GBs for the fcc symmetry.
Figure~\ref{fig:figure3}(a) illustrates the dislocation network forming at the
planar (111) boundary of two twisted fcc crystals. It consists of a hexagonal
network formed by three sets of dislocation directions $\boldsymbol{\xi}_i$
[highlighted in Fig~\ref{fig:figure3}(a) by red dashed lines]. When considering
low angle twist boundaries, such a network made of dislocations with Burgers
vectors $\mathbf{b}=a/2\langle 110 \rangle$, with $a$ the lattice spacing, is
the most favorite according to energy minimization
\cite{Scott1981,DeHosson1990}. This corresponds, e.g., to what has
been observed in twisted, Au (111) thin films
\cite{Schober1969}. For this system, it is known
that a smaller number of dislocation sets with similar Burgers vectors would
be enough to accommodate the misorientation but they would form fourfold
dislocation junctions, instead of the threefold junctions of
Fig~\ref{fig:figure3}(a), and they do not correspond to the energy minimum (see the
green solid line on the dislocation network obtained for $\theta=10^\circ$).
Moreover, this holds true also for the dissociation of the junctions between
dislocations forming additional triangular network \cite{Scott1981}. 
Thus the method used here provides the minimum energy configurations.

Figure~\ref{fig:figure3}(b) illustrates the twist GB obtained at the planar (110)
boundary between two fcc crystals. The resulting dislocation network is
different than for the (111) twist GB. Here it is made by two sub-units, namely
an irregular hexagon and a rhombus. As illustrated in the figure, four sets of
dislocations (see $\boldsymbol{\xi}_i$) are present, still forming three-fold
junctions as for the case reported in Fig~\ref{fig:figure3}(a). For (110)
twist GBs many structures with almost equal energies are known
\cite{Ingle1980,Rittner1996}. However, in analogy with what discussed for the
(111) twist GBs, the configuration in Fig.~\ref{fig:figure3}(b) can be
considered as the most favorite one for small values of $\theta$.

Figure~\ref{fig:figure4} shows the twist GBs for bcc symmetry. In particular,
Fig.~\ref{fig:figure4}(a) shows the dislocation network forming at the planar
(110) boundary between two twisted bcc crystals. An irregular hexagonal structure is
obtained. Such a structure corresponds to what is observed, e.g., for Fe (110) crystals \cite{Ohr1963,Chou1972}. It
consists of two dislocations with Burgers vector $\mathbf{b}_1=a/2\langle 111
\rangle$, along $\boldsymbol{\xi}_1$ and $\boldsymbol{\xi}_2$, which join
forming a dislocation with Burgers vector $\mathbf{b}_2=a \langle 100 \rangle$
along $\boldsymbol{\xi}_3$.  The angle between $\boldsymbol{\xi}_1$ and
$\boldsymbol{\xi}_2$ is $\beta \gtrsim  90^\circ$ \cite{Chou1972}. Similar to
the case reported in Fig.~\ref{fig:figure3}, a possible configuration for
(110) twist GBs for bcc lattice symmetry can consist of two sets of
dislocations oriented along $\langle 112 \rangle$ directions. This structure is
reported by green solid lines in Fig.~\ref{fig:figure4}(a), superposed to the
defects obtained with $\theta=10^\circ$. However, the result of APFC
simulations corresponds to the expected configuration after forming a threefold
junction and represents the more stable configuration. As a further assessment of
the reported results, similar structures have been obtained by MD
\cite{Yang2010} and recently by PFC simulations \cite{Yamanaka2017}. 

Figure~\ref{fig:figure4}(b) illustrates the dislocation network forming at the
(111) planar interface between two twisted crystals. Another dislocation
network is observed here with respect to the (110) twist GB. It is formed by
two sets of hexagons and it is compatible with GBs having six dislocations with
Burgers vector $\mathbf{b}_1$ for the hexagon at the center of the pattern,
surrounded by other six hexagons formed by dislocations with Burgers vectors
$\mathbf{b}_1$ and $\mathbf{b}_2$ as observed in Ref.~\cite{Yamanaka2017} and
illustrated in Fig~\ref{fig:figure4}(b).

\begin{figure}[H]
\center
\includegraphics[width=\linewidth]{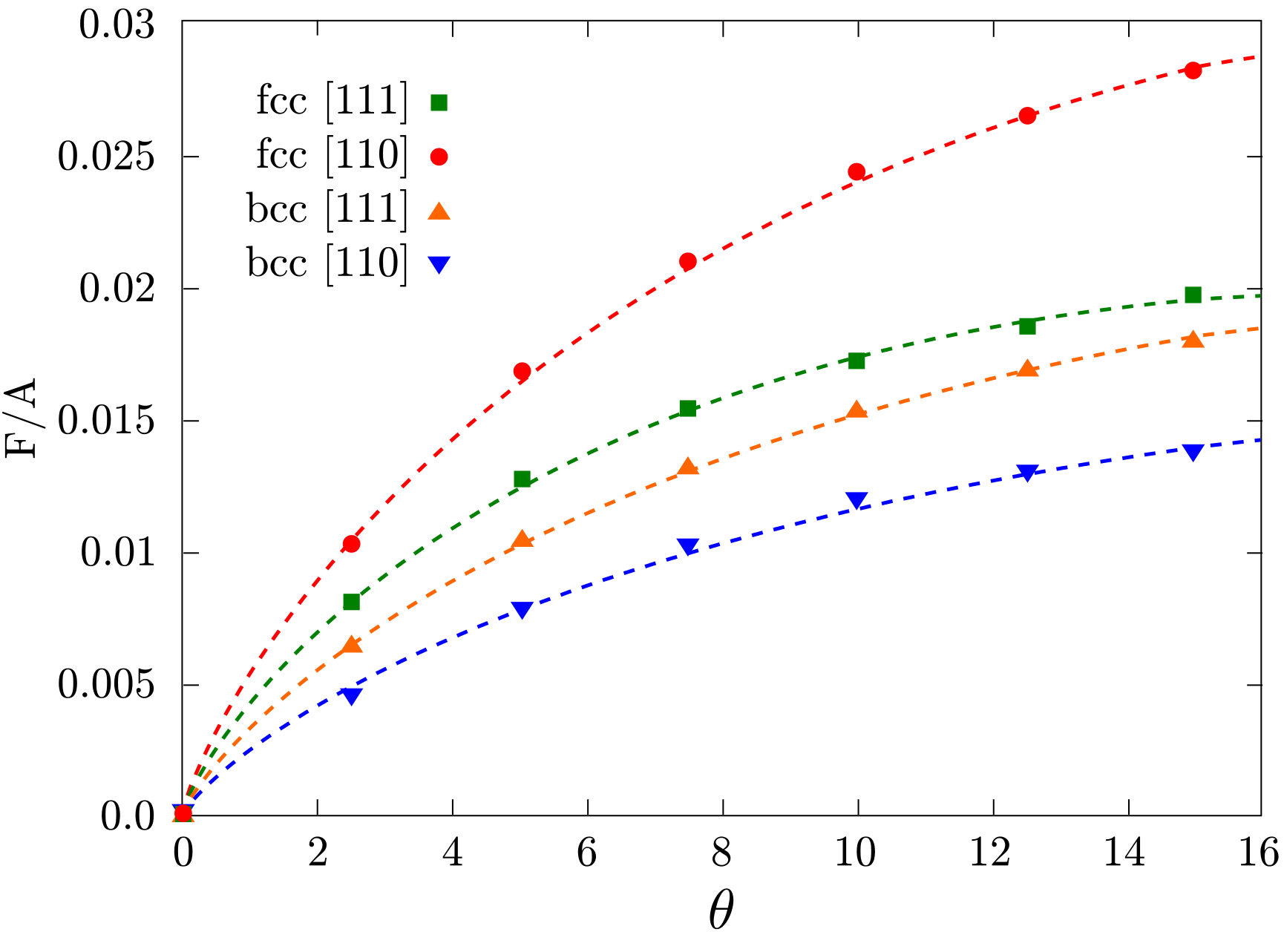} 
    \caption{Energy per unit area (computed by Eq.~\eqref{eq:energyamplitude}, i.e., normalised as in Ref. \cite{ElderPRE2010}) of the twist GBs as a function of the twist
angle. The symmetries and the GB orientations as considered in
Figs.~\ref{fig:figure3} and \ref{fig:figure4} are reported. Dashed lines are
obtained as a fit of the simulation results by Eq.~\eqref{eq:RS}.}
\label{fig:figure5}
\end{figure}

The energies per unit area of the GB computed from
Eq.~\eqref{eq:energyamplitude}, i.e., $F/A$ with $A=2S_yS_z$ (the factor 2 is
included to account for the two GBs formed due to PBCs along the
$\hat{\mathbf{x}}$ direction), as a function of the tilt angle for the symmetries
and the orientations of twist GBs considered so far, are shown in
Fig.~\ref{fig:figure5}. A typical Read-Schockley behavior, namely
corresponding to \begin{equation}
E(\theta) = p \theta[q-\log(\theta)],
\label{eq:RS}
\end{equation}
with $p,q>0$, is observed \cite{Read1950}.
The dashed lines in Fig.~\ref{fig:figure5} are obtained by fitting the energy values by
Eq.~\eqref{eq:RS}. Fcc GBs have higher energy per unit area than the bcc. Focusing on specific symmetries, we
can also see that twist GBs with an orientation corresponding to the plane with
higher packing fractions have lower energies. Indeed, $E^{\rm
fcc}_{(111)}<E^{\rm fcc}_{(110)}$ and $E^{\rm bcc}_{(110)}<E^{\rm
bcc}_{(111)}$. This is in agreement with what was obtained with other methods as,
e.g., in Refs. \cite{Udler1996,Runnels2016b,Runnels2016a}.

\section{Spherical grain-boundaries}
\label{sec:spherical}

\begin{figure*}
\center
\includegraphics[width=\linewidth]{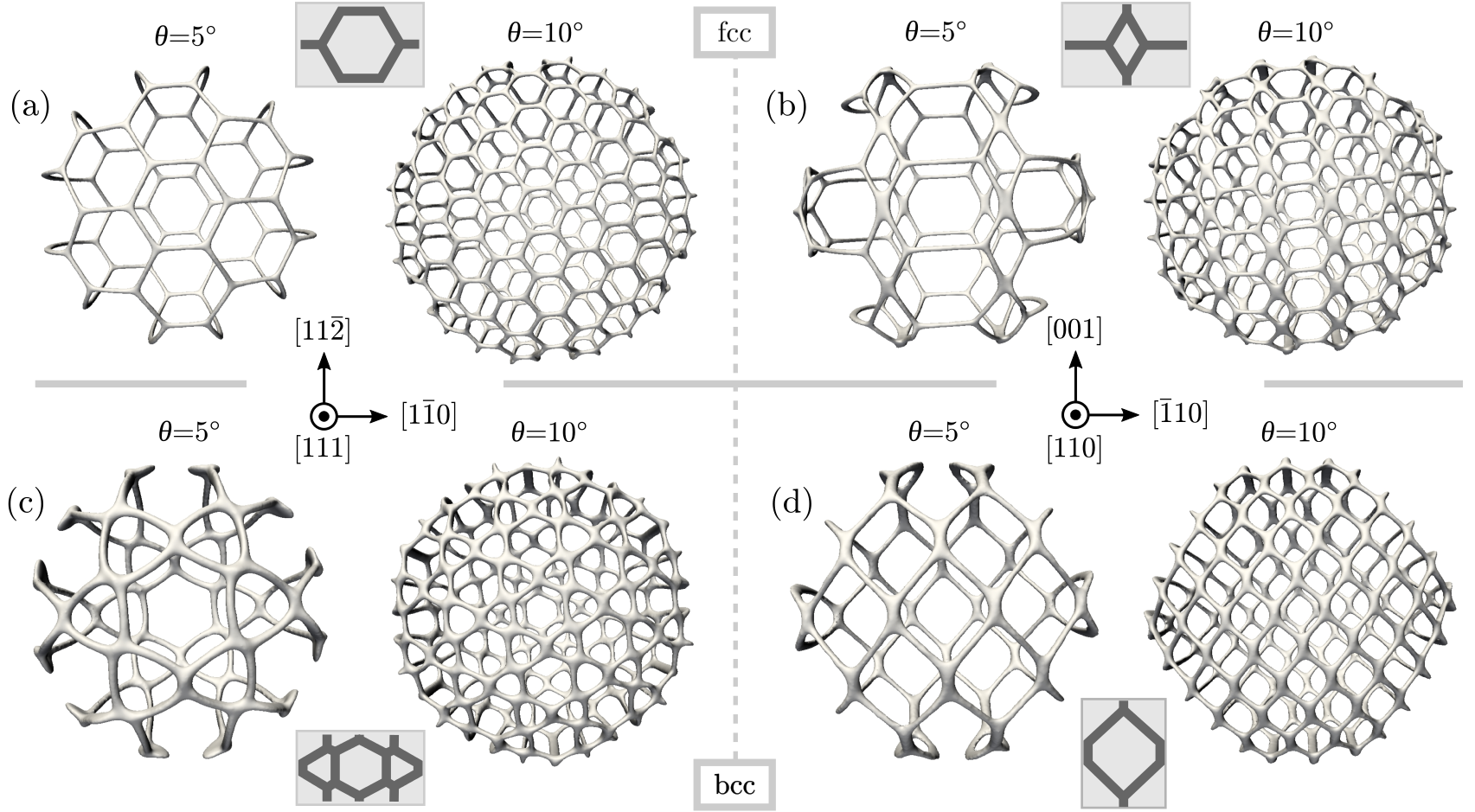} 
    \caption{Dislocation networks forming spherical GBs. A rotated inclusion, with radius of $\sim60\pi$, is
considered in each panel with a rotation of $\theta=5^\circ$ and
$\theta=10^\circ$ about: (a) [111] direction for fcc symmetry, (b) [110]
direction for fcc symmetry, (c) [111] direction for bcc symmetry, (d) [110]
direction for bcc symmetry. The gray insets show the unit-cell of the
dislocation networks found for planar twist GBs and closely resemble the
structure of the spherical GBs where their normal approaches the rotation
axis.} \label{fig:figure6}
\end{figure*}

In this section the APFC approach is exploited to describe the dislocation
networks on spherical GBs, separating a rotated embedded crystal
from an unrotated and relaxed lattice. Moreover, the shrinkage of the resulting
GBs is discussed. The approach illustrated in Sec.~\ref{sec:simdetails}
by means of Figs.~\ref{fig:figure1}(c) and \ref{fig:figure1}(d) is adopted. 

The GB structure generated by a spherical, rotated inclusion in the crystal
structure is expected to be similar to a twist GB in the region where the
surface normal $\hat{\mathbf{n}}$ of the rotated grain approaches the
rotation axis. In directions perpendicular to $\hat{\mathbf{n}}$, pure tilt
GBs are expected while mixed GBs are expected to form for orientations in between.  
The results of APFC simulations, after relaxing to a stationary shape for the 
defect network, are reported in Fig.~\ref{fig:figure6}. 
Fcc and bcc symmetries are considered. In analogy with the investigation illustrated in
Sec.~\ref{sec:planar}, two different rotations about the $[111]$ and $[110]$
directions, both with $\theta=5^\circ$ and $\theta=10^\circ$, are considered.
The unit elements of the dislocation networks obtained for the corresponding
planar, twist GBs are also illustrated by the gray insets with solid lines.

The results reported in Fig.~\ref{fig:figure6} show that the GBs consist of closed dislocation networks. Two regions showing the peculiar networks
corresponding to twist GBs are obtained for $\hat{\mathbf{n}}$ parallel
and antiparallel to the rotation axis. Such networks deform when moving
away from such directions and they are made by elongated straight defects when
approaching directions perpendicular to the rotation axes, typical of pure
tilt GBs. The same qualitative dependence on the rotation angle illustrated previously
for planar twist GBs is found. Notice that, although the limiting case of
twist/tilt GBs can be determined by energy and symmetry considerations and does
not require, in principle, numerical simulations, the mixed GBs forming in
between as well as the connecting dislocation networks are far from trivial and
are obtained here by the same approach delivering planar GBs as in
Sec.~\ref{sec:planar}. Moreover, no restrictions are present on the shape of
the GBs that can be selected arbitrarily. 

\begin{figure*}
\center
\includegraphics[width=\linewidth]{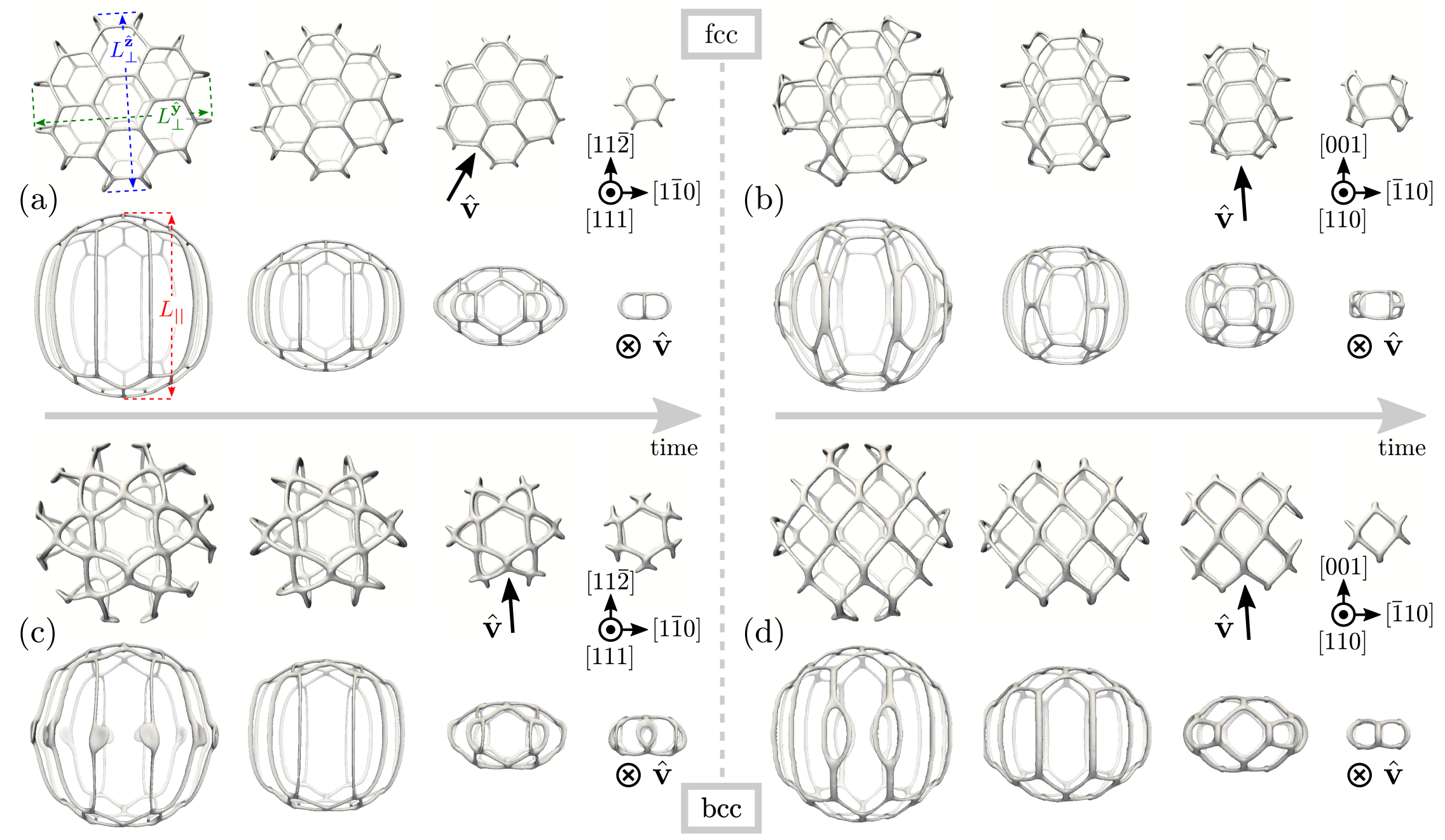} 
    \caption{Shrinkage of spherical GBs for different symmetries with
$\theta=5^\circ$. The views aligned (top) and perpendicular (bottom) to the
rotation axis are shown (as also illustrated by $\hat{\mathbf{v}}$). (a) fcc symmetry, rotation about the [111] direction. (b)
fcc symmetry, rotation about the [110] direction. (c) bcc symmetry, rotation
about the [111] direction. (d) bcc symmetry, rotation about the [110]
direction. Panel (a) illustrates also the width of the embedded crystal along
different directions, namely $L_{||}$, $L_{\perp}^{\hat{\mathbf{y}}}$, and
$L_{\perp}^{\hat{\mathbf{z}}}$.} \label{fig:figure7}
\end{figure*}

The classical model for capillary-driven grain growth predicts a shrinkage of
spherical GBs due to the curvature of the interface between rotated crystals
\cite{Doherty1997}. If an isotropic surface energy is considered the shrinkage
of a spherical grain can be fully described by the decrease of its radius $R$ over
time $t$ such that $R^2=R^2_{\rm ini}-\alpha t$, where $R_{\rm ini}$ is the initial 
radius and $\alpha$ is a constant.  
For real grains, the situation is much more complex, as the dislocation network shrinks 
as a function of time.  Two-dimensional MD and PFC simulations have shown for small-angle 
misorientations that this leads to an increase of the interface energy and a rotation of 
the grain that increases the mismatch orientation \cite{Heinonen2014,Wu2012,Cahn04,Cahn06}.  
This increase in mismatch orientation occurs as the dislocations forming the GB come 
closer together.  Despite these differences the dynamics of the radius (or area) has 
been shown to obey the relationship described above.

The situation is more interesting in three dimensional systems as the anisotropic nature of interface energy and mobility must be taken into account. This implies that the
shrinkage of the grain is more complex as the velocity of the shrinkage depends on the local orientation of the GB. This leads to the presence of
preferential orientation for GBs. An accurate continuum description of shapes originating from anisotropic interface energies and their evolution generally requires advanced methods
\cite{SalvalaglioCGD2015} as well as the proper mapping of GB energies \cite{Kim2014}.
The APFC approach adopted here contains inherently information such as
anisotropic interface energy at GBs (see Fig.~\ref{fig:figure5}) and specific
dislocation structures as discussed in Sec.~\ref{sec:planar}, although it does
not explicitly account for atoms in the crystal lattice.

Long-timescale numerical simulations using the APFC model allow for the investigation of the shrinkage of the
dislocation networks formed at an initially spherical grain.  Figure~\ref{fig:figure7}
illustrates the GB shrinkage achieved for the cases reported in
Fig.~\ref{fig:figure6} with $\theta=5^\circ$. For each case, two views are
shown, illustrating the shrinkage of the grains parallel and perpendicular to
the rotation axis. A common trend independent of the crystal symmetry and orientation appears: The shrinkage in the direction of the rotation axis is faster than in the other directions. Moreover, some small rotations of the grains occur during their shrinkage. In Ref.~\cite{Yamanaka2017} this behavior was obtained (from atomistic calculations
with the PFC model) for the bcc lattice symmetry and attributed to the anisotropic
distribution of dislocations at the GBs. Similar to that work, we also observe 
the same results for the dynamics of the dislocation networks in terms 
of dislocation type and disappearance. This comparison further validates the theoretical and
computational approach adopted here. Moreover, we also observe this qualitative
behavior for the fcc case. 

\begin{figure}
\center
\includegraphics[width=\linewidth]{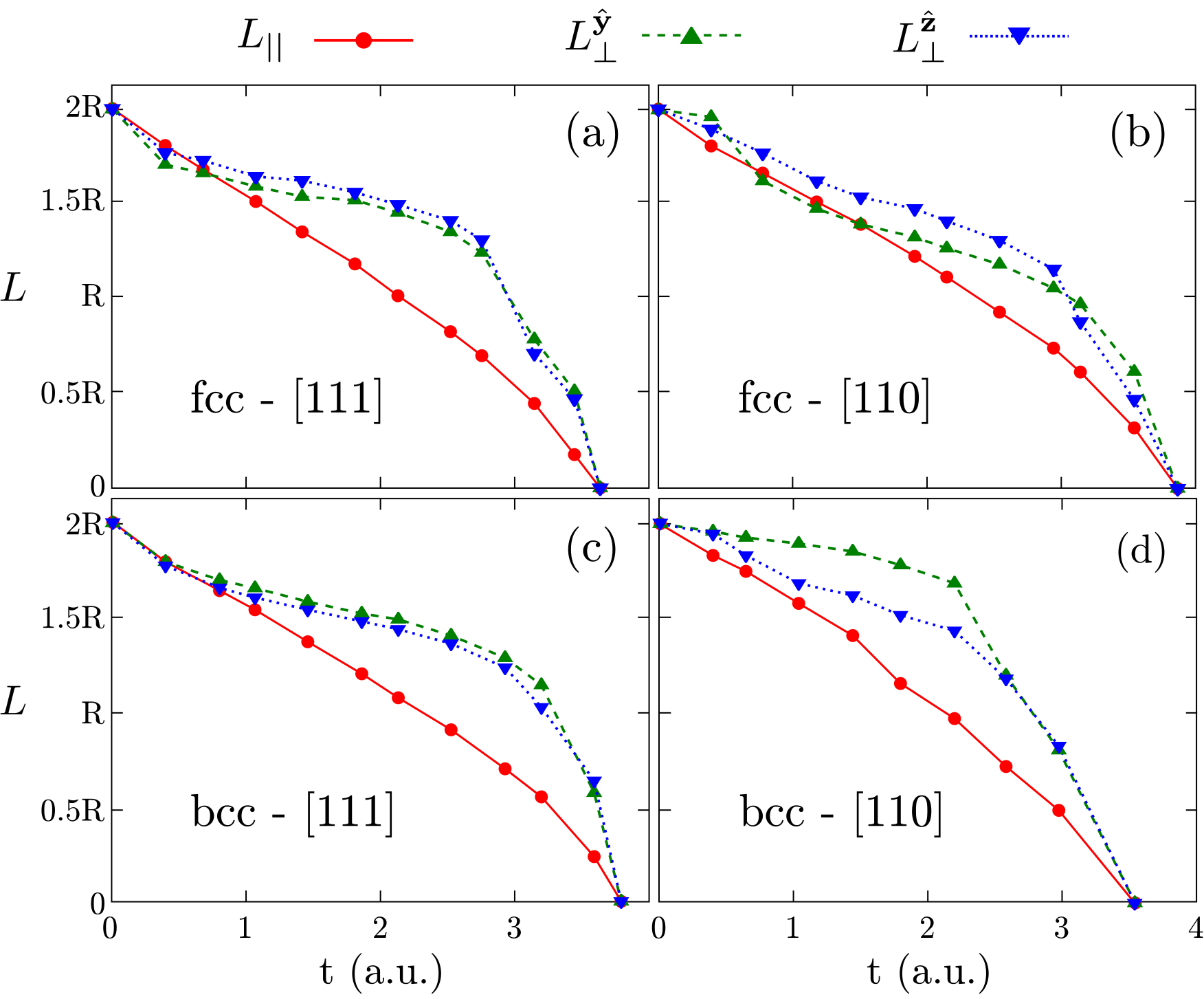} 
    \caption{Widths of the embedded rotated crystal over time, evaluated along
the rotation axis, $L_{||}$, as well as along $\hat{\mathbf{y}}$ and
$\hat{\mathbf{z}}$ of the specific frame of reference, namely
$L_{\perp}^{\hat{\mathbf{y}}}$ and $L_{\perp}^{\hat{\mathbf{z}}}$. The four panels correspond to the cases illustrated
in Fig.~\ref{fig:figure7}.} \label{fig:figure8}
\end{figure}

Deeper insights on the anisotropic shrinkage can be found in Fig.~\ref{fig:figure8}. Here, the
widths $L$ of the shrinking GBs are extracted along the rotation axis
$L_{||}$ and along two directions in the plane perpendicular to the rotation
axis, $L_{\perp}^{\hat{\mathbf{y}}}$ and $L_{\perp}^{\hat{\mathbf{z}}}$ namely
evaluated along the $\hat{\mathbf{y}}$ and $\hat{\mathbf{z}}$ directions of
different frames of reference as specified in Sec.~\ref{sec:simdetails} (see
also Fig.~\ref{fig:figure7}). As seen in Fig.~\ref{fig:figure8} an initial transient phase can be 
observed corresponding to the relaxation of the initial condition and the
formation of defect networks. Afterwards, a smaller width along the rotation
axis ($L_{||}$) is observed for all the cases. At the end of the process when all the
widths approach zero, a slightly faster evolution is observed. This corresponds
to the final annihilation of the dislocations and the disappearance of the
grains. $L_{||}$ is found to decrease linearly with nearly the same velocity
during the entire process. Conversely, after the formation of the dislocation
networks, both $L_{\perp}^{\hat{\mathbf{y}}}$ and
$L_{\perp}^{\hat{\mathbf{z}}}$ show an almost linear behavior as well but with a velocity
significantly slower than $L_{||}$. This stage can be ascribed to a stronger tendency to shrink the two-dimensional dislocation networks having a normal perpendicular to the rotational axis rather than the dislocation network corresponding to twist GBs. When approaching the final
disappearance of the grain, a sudden increase of the velocity along $\hat{\mathbf{y}}$ and $\hat{\mathbf{z}}$ directions is observed. This change in the shrinking rate, in the presence of an almost constant decrease of $L_{||}$, allows for having a closed dislocation network preventing the formation of a high energy configuration such as a platelike (2D) embedded crystal. Indeed, this would exhibit too large curvatures of the GB, that would correspond to a too high interfacial energy between the rotated inclusion and the unrotated crystal. Notice that, although a good description of the defects is achieved, the position of atoms during the annihilation of defects are not expected to be carefully described by the APFC approach.

\begin{figure}[H]
\center
\includegraphics[width=\linewidth]{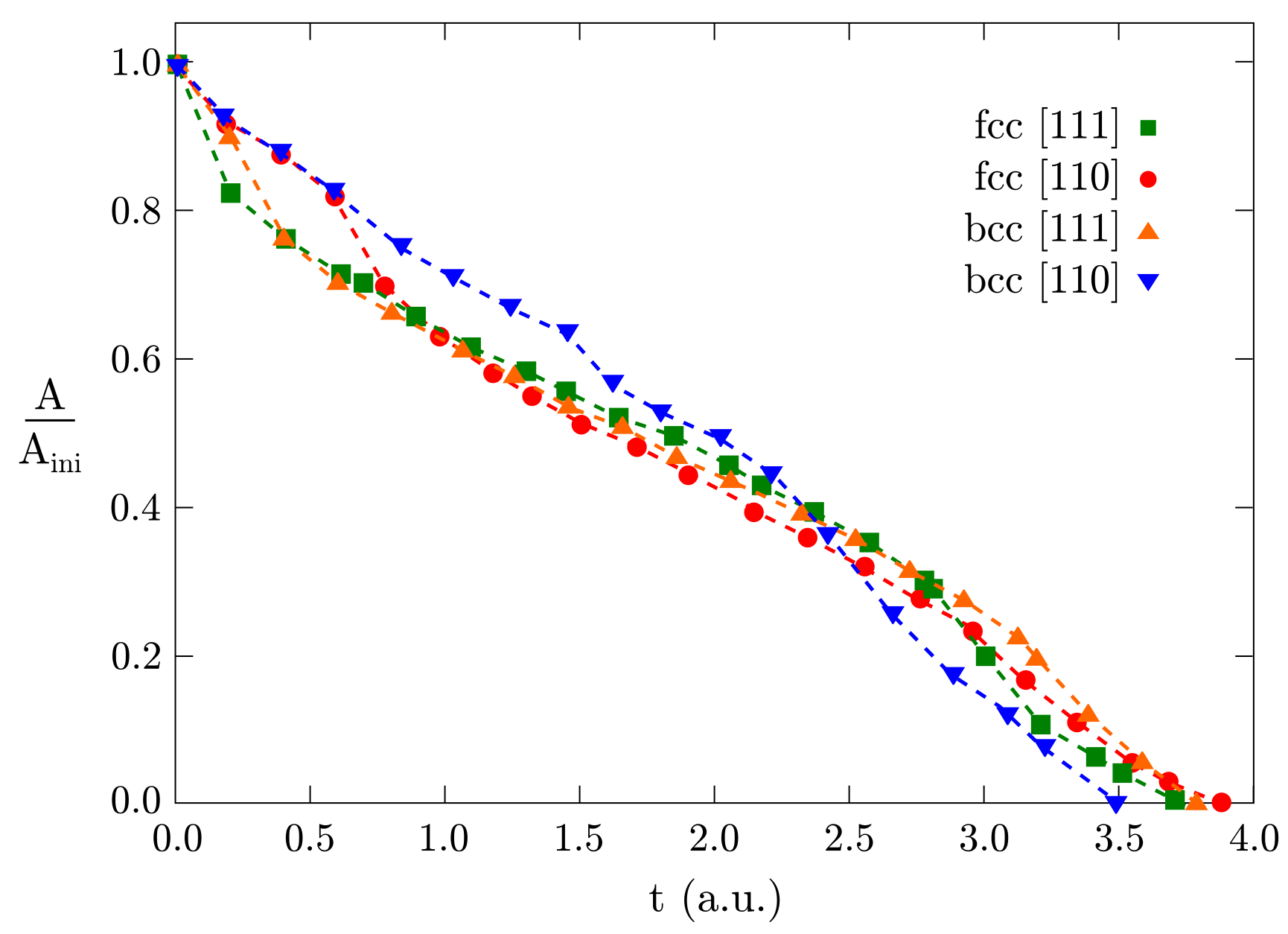} \caption{Surface area of the
shrinking crystals of Fig.~\ref{fig:figure7} approximated as an ellipse
having $L_{||}$, $L_{\perp}^{\hat{\mathbf{y}}}$, and
$L_{\perp}^{\hat{\mathbf{z}}}$ as axes.}
    \label{fig:figure9}
\end{figure}

 In Ref.~\cite{Yamanaka2017} a linear scaling
of the GBs area during the evolution has been also found recovering the
prediction of the classical theory \cite{Doherty1997}. From the data in Fig.~\ref{fig:figure8}, the
surface area of the grains can be computed 
by assuming an ellipsoidal shape where the width of the grains is represented by its
axes. In particular, the surface area can be approximated by the Knud-Thomsen formula,
\begin{equation}
A=\frac{\pi}{3^{1/p}}\left[\left(L_{||}L_{\perp}^{\hat{\mathbf{y}}}\right)^p+\left(L_{||}L_{\perp}^{\hat{\mathbf{z}}}\right)^p
+\left(L_{\perp}^{\hat{\mathbf{y}}} L_{\perp}^{\hat{\mathbf{z}}} \right)^p\right]^{1/p},
\end{equation}
with $p=1.6075$. The corresponding areas normalized by the initial surface of
the embedded spherical grain $A_{\rm ini}=4\pi R^2$ with $R=60\pi$ are reported
in Fig.~\ref{fig:figure9}.  
The decrease of the surface area
of the grains over time is nearly linear for
all the considered cases. Small deviations are observed only in the first and
last stages corresponding to the defect formation and to the final
disappearance of the grain. Moreover, from both Figs.~\ref{fig:figure8} and
\ref{fig:figure9} one can notice that the duration of the entire shrinkage
process is similar even considering different symmetries and orientation of the
GB. This evidence is in agreement with the results reported by PFC simulations
in Ref.~\cite{Yamanaka2017} concerning bcc lattice with different rotation
axes. The dynamics of fcc rotated grains are then found to follow a very
similar kinetic pathway, characterized by an anisotropic shrinkage with smaller
$L_{||}$ and a similar timescale towards the disappearance of the rotated
inclusion.

It is worth mentioning that the considered APFC approach does not include the effects of the dynamics mediated by phonons. Such a contribution may play a role when accounting for fast coarsening dynamics, providing correction to the shrinking rate which may be dependent on the lattice symmetry\cite{Heinonen2016}. However, we expect the main qualitative observations to remain unaltered also in this regime, while conclusion reached by this study still hold true quantitatively for relatively slow processes.

\section{Conclusions}
\label{sec:conclusions}
In this paper, we illustrated how the APFC approach can be
adopted to provide a detailed coarse-grained, three-dimensional description of
dislocation networks at GBs. Their morphology and evolution have been simulated
by means of the evaluation of fields (namely, the complex amplitudes $\eta_j$) varying on larger lengthscales than the lattice spacing. Thus, a spatial resolution typically larger than the one used in atomistic approaches can be used to
provide information about GBs on large systems and long timescales. Moreover,
the extended defects forming at GBs can be directly identified by a scalar
order parameter, $A^2$, which can be also exploited to modify and control defect
properties \cite{SalvalaglioAPFC2017}. The method has been shown to tackle fcc
and bcc lattice symmetries without major changes. Different rotations and GBs
orientations can be also considered without any restriction as shown by the
reported simulations. 

The results on planar, twist GBs provided an assessment of the 
validity of the approach used in this work for obtaining 
GB morphologies.
In particular, the fine structures made of
dislocations were shown to be directly accessible and 
consistent with well-known results obtained by symmetry considerations, energy
minimization, and other simulation methods. Notice that the relevant
case of pure-twist GBs has been addressed but no restrictions are present for
the simulation of planar, pure-tilt, or mixed GBs. The dislocation networks
obtained in the case of spherical GBs show the possibility to account for
arbitrary GBs geometries, without loss of descriptive capability and accuracy.

The dynamics of dislocation networks at GBs during the shrinkage of embedded
rotated crystals was also addressed. The importance of this investigation
is twofold. First, the dynamics of defects achieved by APFC simulation is
found to reproduce recent results obtained by PFC simulations
accounting for bcc lattice symmetries \cite{Yamanaka2017}, thus further validating the general APFC approach as a
coarse-grained description of the PFC model. In particular, the peculiar
anisotropic shrinkage of spherical rotated grains is reproduced with a linear
decrease of the interface between rotated and unrotated crystal. Second, the
approach allows for simulating and analyzing grain shrinkage for fcc
symmetries. The dynamics of spherical GBs in fcc crystals is found to occur
with very similar features with respect to bcc crystals.
Specifically, the feature of anisotropic shrinkage
observed for bcc both here and in
Ref.~\cite{Yamanaka2017} is then found to be general and independent of the
lattice symmetry as well as of the rotation axis, even
accounting for the different dislocation networks forming in each case. This
unveils the generality of the observed behaviors and deepens the knowledge
about GBs dynamics. 

This work paves further the way to detailed investigations of 3D systems by
APFC including also other physical contributions already shown to be correctly
modeled by this approach in 2D such as binary systems
\cite{ElderPRE2010,ElderJPCM2010} or the presence of both GBs and compositional
domains \cite{Geslin2015,Xu2016}. Further work will be devoted to the
optimization of the numerical approach in order to provide even more efficient
calculations to enable simulations of grain growth in polycrystalline materials, similar to Ref.~\cite{Backofen2014} in 2D.

\section*{Acknowledgements}
M.S. acknowledges the support of the Postdoctoral Research Fellowship awarded
by the Alexander von Humboldt Foundation.  R.B. and A.V. acknowledge the
financial support from the German Research Foundation (DFG) under Grant No. SPP
1959. K.R.E. acknowledges financial support from the National Science
Foundation under Grant No. DMR1506634. The computational resources were
provided by ZIH at TU Dresden and by the J\"ulich Supercomputing Center within
the Project No. HDR06.

\end{document}